\documentclass[12pt]{article}
\usepackage{amssymb,amsmath,epsfig}

\begin{document}

\title{\bf Effects of $f(R)$ Model on the Dynamical Instability of Expansionfree Gravitational Collapse}

\author{M. Sharif \thanks{msharif.math@pu.edu.pk} and H. Rizwana Kausar
\thanks{rizwa\_math@yahoo.com}\\
Department of Mathematics, University of the Punjab,\\
Quaid-e-Azam Campus, Lahore-54590, Pakistan.}

\date{}
\maketitle

\begin{abstract}
Dark energy models based on $f(R)$ theory have been extensively
studied in literature to realize the late time acceleration. In
this paper, we have chosen a viable $f(R)$ model and discussed its
effects on the dynamical instability of expansionfree fluid
evolution generating a central vacuum cavity. For this purpose,
contracted Bianchi identities are obtained for both the usual
matter as well as dark source. The term dark source is named to
the higher order curvature corrections arising from $f(R)$
gravity. The perturbation scheme is applied and different terms
belonging to Newtonian and post Newtonian regimes are identified.
It is found that instability range of expansionfree fluid on
external boundary as well as on internal vacuum cavity is
independent of adiabatic index $\Gamma$ but depends upon the
density profile, pressure anisotropy and $f(R)$ model.
\end{abstract}
{\bf Keywords:} $R+\delta R^2$ gravity; Instability; Expansionfree evolution.\\
{\bf PACS:} 04.50.Kd

\section{Introduction}

The "modified gravity" has become a standard terminology for the
theories describing gravitational interactions which differ from
the most conventional theory of general relativity (GR). In these
modified theories, $f(R)$ gravity is able to mimic the standard
$\Lambda$CDM cosmological evolution and dark energy (DE) problem.
Since the laws of gravity gets modified on large distances in
$f(R)$ models, this leaves several interesting observational
signatures such as modification to the spectra of the galaxy
clustering \cite{cb}, cosmic microwave background \cite{zs} and
weak lensing \cite{st}. These models have some vacuum solutions
with null scalar curvature that allow to recover certain GR
solutions. In addition, it has many other applications such as
inflation, local gravity constraints, cosmological perturbations
and spherically symmetric solutions in weak and strong
gravitational backgrounds.

The most important feature of $f(R)$ gravity is to provide the
very natural gravitational alternative for DE without adding any
matter component. Let us now show that how it can be related to
the problem of DE by a straightforward argument. When the
Einstein-Hilbert (EH) gravitational action in GR,
\begin{equation}\label{a}
S_{EH}=\frac{1}{2 \kappa}\int d^{4}x\sqrt{-g}R,
\end{equation}
is written in the modified form as follows
\begin{equation}\label{b}
S_{f(R)}=\frac{1}{2\kappa}\int d^{4}x\sqrt{-g}f(R),
\end{equation}
the addition of a non-linear function of the Ricci scalar
demonstrates to cause acceleration for a wide variety of $f(R)$
function, e.g., \cite{fr1}-\cite{fr10}. Variation of $f(R)$ action
with respect to the metric tensor leads to the following fourth
order partial differential equations
\begin{equation}\label{b'}
F(R)R_{\alpha\beta}-\frac{1}{2}f(R)g_{\alpha\beta}-\nabla_{\alpha}
\nabla_{\beta}F(R)+ g_{\alpha\beta} \Box F(R)=\kappa
T_{\alpha\beta},\quad(\alpha,\beta=0,1,2,3),
\end{equation}
where $F(R)\equiv df(R)/dR$. Writing this equation in the form of
Einstein tensor, it follows that \cite{fr11}
\begin{equation}\label{12}
G_{\alpha\beta}=\frac{\kappa}{F}(T_{\alpha\beta}^{m}+T_{\alpha\beta}^{(D)}),
\end{equation}
where
\begin{equation}\label{d}
T_{\alpha\beta}^{(D)}=\frac{1}{\kappa}\left[\frac{f(R)-R
F(R)}{2}g_{\alpha\beta}+\nabla_{\alpha} \nabla_{\beta}F(R)
-g_{\alpha\beta} \Box F(R)\right].
\end{equation}
Equation (\ref{12}) shows that effective stress-energy tensor
$T_{\alpha\beta}^{(D)}$ plays the role of matter source in the
field equations with purely geometrical origin. This approach may
provide all the matter ingredients needed to tackle the dark side
of the universe. Thus if we restrict "dark source" such that it
does not satisfy the usual energy conditions then it can play the
role of both dark matter and DE. Consequently, this theory may be
used to explain the expansion of the universe and effects of DE in
gravitational phenomena. This approach is sometimes convenient to
use when we study the DE equation of state \cite{hs} as well as
the equilibrium description of thermodynamics for the horizon
entropy \cite{Ba}. The trace of Eq.(\ref{b'}) determines the
dynamics of the scalar field $F(R)$ given by
\begin{equation}\label{b''}
F(R)R-2f(R)+3\Box F(R)=\kappa T.
\end{equation}
Notice that there exists a de Sitter point which corresponds to a
vacuum solution ($T=0$) at which the Ricci scalar is constant. Since
$\Box F(R)=0$ at this point, so we obtain from the above equation
\begin{equation}\label{b'''}
F(R)R-2f(R)=0.
\end{equation}
It is mentioned here that the model $f(R)=\delta R^2$ satisfies this
condition and yields the exact de Sitter solution \cite{staro}. In
the model $f(R)=R+\delta R^2$, because of the linear term in $R$,
the inflationary expansion ends when $\delta R^2$ becomes smaller
than the linear term. This is followed by a reheating stage in which
the oscillation of $R$ leads to the gravitational particle
production. It is also possible to use the de Sitter point for DE.

The gravitational collapse is an important and long standing issue
in GR. Recently, it has gained attention in modified theories as
well. During gravitational collapse, self-gravitating objects may
pass through phases of intense dynamical activities for which
quasi-static approximation is not reliable. For instance, the
collapse of very massive stars \cite{ben}, the quick collapse
phase yielding neutron star formation \cite{myra} and the peculiar
stars. The dynamical equations are used to observe the collapsing
process while the vanishing expansion scalar condition is
developed in connection with the description of voids. The
vanishing of expansion scalar requires that the innermost shell of
the fluid should be away from the center, initiating therefrom the
formation of the cavity within the matter distribution
\cite{H1,H2}. This natural appearance of a vacuum cavity suggests
that they might be relevant for the moddelling of cosmological
voids. The Skripkin model is the first example satisfying
expansionfree condition \cite{skrip}. This model corresponds to
evolution of spherically symmetric non-dissipating fluid
distribution with constant energy density. It is also remarked
that the expansionfree condition is a sufficient but not a
necessary condition for the appearance of cavities. Cavities
described under different kinematical condition are discussed by
Herrera et al \cite{Hcqg}.

For the physically relevant models, the expansionfree evolution
requires pressure anisotropy in the fluid distribution and
inhomogeneity in the energy density. To assume isotropy of the
pressure together with the expansion-free condition, we impose
that the energy-density is independent on the time-like
coordinate, which severely restricts the models \cite{Hprd79}. A
stellar model can exist only if it is stable against fluctuations.
The problem of dynamical instability is closely related to the
structure formation and evolution of self-gravitating objects.
Chandrasekhar \cite{chand} was the first who worked in this
direction. Afterwards, this issue has been investigated by many
authors for adiabatic, non-adiabatic, anisotropic and shearing
viscous fluids \cite{H3}-\cite{H4}. We have investigated the
problem of DE and gravitational collapse in $f(R)$ gravity
\cite{SR1,SR2}.

In this paper, we are concerned with spherically symmetric stars
having locally anisotropic fluid distribution inside and would
investigate how $f(R)$ terms affect the dynamical instability of
expansionfree fluid evolution. The format of the paper is as
follows. In section \textbf{2}, basic equations are given. Section
\textbf{3} is devoted to study the perturbation scheme and a
well-known physical $f(R)$ model. In section \textbf{4}, Newtonian
and post Newtonian approximations are taken into account and
different terms belonging to these regimes are identified. Also,
dynamical equations are investigated under the conditions of
vanishing scalar and instability conditions of fluid evolution are
discussed. The last section \textbf{5} provides the summary of the
work.

\section{Field Equations and Dynamical Equations}

We consider a $3D$ hypersurface $\Sigma^{(e)}$, an external
boundary of the collapsing spherically symmetric star, which
divides a $4D$ spacetime into two regions named as interior and
exterior spacetimes. The interior spacetime to $\Sigma^{(e)}$ is
described by the most general spherically symmetric metric as
follows
\begin{equation}\setcounter{equation}{1} \label{1}
ds^2_-=A^2(t,r)dt^{2}-B^2(t,r)dr^{2}-C^2(t,r)(d\theta^{2}+\sin^2\theta
d\phi^{2}).
\end{equation}
The exterior spacetime is described by the Schwarzschild metric
given by
\begin{equation}\label{25}
ds^2_+=\left(1-\frac{2M}{r}\right)d\nu^2+2drd\nu-r^2(d\theta^2+\sin^2\theta
d\phi^2),
\end{equation}
where $M$ represents the total mass of the system inside the
boundary surface $\Sigma^{(e)}$ and $\nu$ is the retarded time.
The fluid filling the spherically symmetric star is assumed to be
locally anisotropic with inhomogeneous energy density. Thus the
energy-momentum tensor for such a fluid is given by
\begin{equation}\label{2}
T_{\alpha\beta}=(\rho+p_{\perp})u_{\alpha}u_{\beta}-p_{\perp}g_{\alpha\beta}+
(p_r-p_{\perp})\chi_{\alpha} \chi_{\beta},
\end{equation}
where $\rho$ is the energy density, $p_{\perp}$ the tangential
pressure, $p_r$ the radial pressure, $u_{\alpha}$ the
four-velocity of the fluid and $\chi_{\alpha}$ is the unit
four-vector along the radial direction. These quantities satisfy
the relations
\begin{equation}\label{3}
u^{\alpha}u_{\alpha}=1,\quad\chi^{\alpha}\chi_{\alpha}=-1,\quad
\chi^{\alpha}u_{\alpha}=0
\end{equation}
and are obtained from the following definitions in co-moving
coordinates
\begin{equation}\label{6}
u^{\alpha}=A^{-1}\delta^{\alpha}_{0},\quad
\chi^{\alpha}=B^{-1}\delta^{\alpha}_{1}.
\end{equation}
The expansion scalar, $\Theta$, is given by
\begin{equation}\label{8}
\Theta=u^{\alpha}_{;\alpha}=\frac{1}{A}\left(\frac{\dot{B}}{B}+2\frac{\dot{C}}{C
}\right),
\end{equation}
where dot and prime represent derivatives with respect to $t$ and
$r$ respectively. For spherically symmetric interior metric, the
field equations (\ref{12}) become
\begin{eqnarray}\nonumber
&&\left(\frac{2\dot{B}}{B}+\frac{\dot{C}}{C}\right)
\frac{\dot{C}}{C}-\left(\frac{A}{B}\right)^2
\left[\frac{2C''}{C}+\left(\frac{C'}{C}\right)^2-\frac{2B'C'}{BC}
-\left(\frac{B}{C}\right)^2\right]=\frac{\kappa}{F}\left[{\rho}A^{2}\right.\\\label{13}
&&\left.+\frac{A^2}{\kappa}\left\{\frac{f-R
F}{2}+\frac{F''}{B^2}+\left(\frac{2\dot{C}}{C}-\frac{\dot{B}}{B}\right)
\frac{\dot{F}}{A^2}+\left(\frac{2C'}{C}-\frac{B'}{B}\right)\frac{F'}{B^2}\right\}\right],
\\\label{14}
&&-2\left(\frac{\dot{C'}}{C}-\frac{\dot{C}A'}{CA}-\frac{\dot{B}C'}{BC}\right)
=\frac{1}{F}\left(\dot{F}'
-\frac{A'}{A}\dot{F}-\frac{\dot{B}}{B}F'\right),\\\nonumber
&&-\left(\frac{B}{A}\right)^2\left[\frac{2\ddot{C}}{C}-\left(\frac{2\dot{A}}{A}
-\frac{\dot{C}}{C}\right) \frac{\dot{C}}{C}\right]
+\left(\frac{2A'}{A}+\frac{C'}{C}\right)\frac{C'}{C}-\left(\frac{B}{C}\right)^2
=\frac{\kappa}{F}
\left[p_rB^{2}\right.
\\\label{15}
&&\left.-\frac{B^2}{\kappa}\left\{\frac{f-R
F}{2}-\frac{\ddot{F}}{A^2}+\left(\frac{\dot{A}}{A}+\frac{2\dot{C}}{C}\right)
\frac{\dot{F}}{A^2}+\left(\frac{A'}{A}+\frac{2C'}{C}\right)\frac{F'}{B^2}\right\}\right],\\\nonumber
&&-\left(\frac{C}{A}\right)^2\left[\frac{\ddot{B}}{B}-\frac{\ddot{C}}{C}-\frac{\dot{A}}{A}
\left(\frac{\dot{B}}{B}+\frac{\dot{C}}{C}\right)+\frac{\dot{B}\dot{C}}{BC}\right]
+\left(\frac{C}{B}\right)^2\left[\frac{A''}{A}+\frac{C''}{C}-\frac{A'B'}{AB}\right.\\\nonumber
&&\left.+\left(\frac{A'}{A}-\frac{B'}{B}\right)\frac{C'}{C}\right]
=\frac{\kappa}{F}\left[p_{\perp}C^2-\frac{C^2}{\kappa}\left\{\frac{f-R
F}{2}-\frac{\ddot{F}}{A^2}+\frac{F''}{B^2}+
\frac{\dot{F}}{A^2}\right.\right.\\\label{16}
&&\times\left.\left.\left(\frac{\dot{A}}{A}-\frac{\dot{B}}{B}+\frac{\dot{C}}{C}\right)
+\left(\frac{A'}{A}-\frac{B'}{B}+\frac{C'}{C}\right)\frac{F'}{B^2}\right\}\right].
\end{eqnarray}

The dynamical equations help to study the properties of collapsing
process. To formulate these equations, the mass function
introduced by Misner and Sharp is defined as follows \cite{MS}
\begin{equation}\label{18}
m(t,r)=\frac{C}{2}(1+g^{\mu\nu}C_{,\mu}C_{,\nu})=\frac{C}{2}\left(1+\frac{\dot{C}^2}{A^2}
-\frac{C'^2}{B^2}\right).
\end{equation}
This equation provides the total energy inside a spherical body of
radius $"C"$. From the continuity of the first and second
differential forms, the matching of the adiabatic sphere to the
Schwarzschild spacetime on the boundary surface, ${\Sigma^{(e)}}$,
yields the following result \cite{H2,Rchan}
\begin{equation}\label{28}
M\overset{\Sigma^{(e)}}{=}m(t,r)
\end{equation}
and
\begin{eqnarray}\nonumber
&&2\left(\frac{\dot{C'}}{C}-\frac{\dot{C}A'}{CA}-\frac{\dot{B}C'}{BC}\right)
\overset{\Sigma^{(e)}}{=}
-\frac{B}{A}\left[\frac{2\ddot{C}}{C}-\left(\frac{2\dot{A}}{A}-
\frac{\dot{C}}{C}\right) \frac{\dot{C}}{C}\right]\\\label{29} &&
+\frac{A}{B}\left[\left(\frac{2A'}{A}+\frac{C'}{C}\right)\frac{C'}{C}
-\left(\frac{B}{C}\right)^2\right].
\end{eqnarray}
Using the field equations (\ref{14}) and (\ref{15}) in the above
equation, we obtain
\begin{equation}\label{30}
-p_r\overset{\Sigma^{(e)}}{=}\frac{T^{(D)}_{11}}{B^2}
+\frac{T^{(D)}_{01}}{AB}.
\end{equation}

As the expansion scalar describes the rate of change of small
volumes of the fluid. It is interesting to note that expansionfree
models define the two hypersurfaces, one separating the fluid
distribution externally from the Schwarzschild vacuum solution
while the other is boundary between the central Minkowskian cavity
and the fluid. Taking ${\Sigma^{(i)}}$ ($i$ stands for internal)
to be the boundary surface of that vacuum cavity and matching it
with Minkowski spacetime, we have
\begin{equation}\label{31}
m(t,r)\overset{\Sigma^{(i)}}{=}0, \quad
-p_r\overset{\Sigma^{(i)}}{=}\frac{T^{(D)}_{11}}{B^2}
+\frac{T^{(D)}_{01}}{AB}.
\end{equation}
The physical applications of such models lie at the core of
astrophysical background where the cavity within the fluid
distribution is present. It may help to investigate the voids on
cosmological scales \cite{liddle}. Voids are the spongelike
structures and occupying 40\%-50\% volume of the universe. There
exist different sizes of the voids, i.e., mini-voids \cite{AV} to
super-voids \cite{LR}. Observational data shows that voids are
neither empty nor spherical. However, for the sake of
investigations they are usually described as vacuum spherical
cavities surrounding by the fluid.

The proper time and radial derivatives are given by
\begin{equation}\label{22}
D_{T}=\frac{1}{A}\frac{\partial}{\partial t},\quad
D_{C}=\frac{1}{C'}\frac{\partial}{\partial r},
\end{equation}
where $C$ is the areal radius of the spherical surface. The
velocity of the collapsing fluid is defined by the proper time
derivative of $C$, i.e.,
\begin{equation}\label{23}
U=D_{T}C=\frac{\dot{C}}{A}
\end{equation}
which is always negative. Using this expression, Eq.(\ref{18})
implies that
\begin{equation}\label{23'}
E\equiv\frac{C'}{B}=\left[1+U^{2}+\frac{2m}{C}\right]^{1/2}.
\end{equation}
The rate of change of mass in Eq.(\ref{18}) with respect to proper
time, with the use of Eqs.(\ref{13})-(\ref{16}), is given by
\begin{equation}\label{25'}
D_{T}m=\frac{-\kappa}{2F}\left[\left(p_r+\frac{T^{(D)}_{11}}{B^2}\right)U
-E\frac{T^{(D)}_{01}}{AB}\right]C^2.
\end{equation}
This represents variation of total energy inside a collapsing
surface of radius $C$. The presence of dark fluid components shows
the contribution of DE having large negative pressure. These terms
appear with positive sign representing negative effect, hence
decrease the rate of change of mass with respect to time. Now, it
depends upon the strength of DE terms that they may balance the
positive effect of effective radial pressure or overcome on them.
Likewise, we have
\begin{equation}\label{26}
D_{C}m=\frac{\kappa}{2F}\left[\rho+\frac{T^{(D)}_{00}}{A^2}
-\frac{U}{E}\frac{T^{(D)}_{01}}{AB}\right]C^2.
\end{equation}
This equation describes how energy density and curvature terms
influence the mass between neighboring surfaces of radius $C$ in
the fluid distribution. Here the rate would decrease in the
consecutive surfaces by the repulsive effect of DE. Integration of
Eq.(\ref{26}) with respect to "$C$" leads to
\begin{equation}\label{24}
m=\kappa\int^{C}_{0}\frac{C^2}{2F}\left[\rho+\frac{T^{(D)}_{00}}{A^2}
-\frac{U}{E}\frac{T^{(D)}_{01}}{AB}\right]dC.
\end{equation}

The dynamical equations can be obtained from the contracted
Bianchi identities. Consider the following two equations
\begin{eqnarray}\label{bb}
\left(T^{\alpha\beta}+\overset{(D)}{T^{\alpha\beta}}\right)_{;\beta}u_{\alpha}=0,\quad
\left(T^{\alpha\beta}+\overset{(D)}{T^{\alpha\beta}}\right)_{;\beta}
\chi_{\alpha}=0
\end{eqnarray}
which yield respectively
\begin{eqnarray}\label{19}
&&\frac{\dot{\rho}}{A}+\frac{\dot{B}}{AB}(\rho+p_r)+
\frac{2}{A}\frac{\dot{C}}{C}(\rho+p_{\perp})+D_1=0,\\\label{21}
&&\frac{p_r'}{B}+(\rho+p_r)\frac{A'}{AB}+2(p_r-p_{\perp})\frac{C'}{BC}
+D_2 =0,
\end{eqnarray}
where $D_1$ and $D_2$ are components of dark source given in
\textbf{Appendix} (\ref{19*}, \ref{21*}).

\section{The $f(R)$ Model and Perturbation Scheme}

Obtaining the correct dynamics of the background cosmological model
is not sufficient for any theory to be viable. It is practically
impossible to separate different $f(R)$ models without using
cosmological perturbations. In this section, we apply perturbation
scheme on the field equation and dynamical equations in order to
investigate the instability conditions of collapsing fluid
evolution. We consider following $f(R)$ model
\begin{equation}\setcounter{equation}{1}\label{fr}
f(R)=R+\delta R^2,
\end{equation}
where $\delta$ is any real number. When we take conformal
transformation of $f(R)$ action, it is found that the
Schwarzschild solution is the only static spherically symmetric
solution for the above model \cite{whitt}. The stability criteria
for this model is restricted to $\delta>0$ which corresponds to
$f^{''}(R)>0$. For $\delta=0$, GR is recovered in which black
holes are stable classically but not quantum mechanically due to
Hawking radiations. Since such features also found in $f(R)$
gravity, hence the classical stability condition for the
Schwarzschild black hole can be expressed as $f^{''}(R)>0$
\cite{sot}.

In our perturbation scheme, we assume that initially all the
quantities have only radial dependence, i.e., fluid is in static
equilibrium. After that all the quantities and the metric
functions have time dependence in their perturbation. This is
given by
\begin{eqnarray}\label{41}
A(t,r)&=&A_0(r)+\epsilon T(t)a(r),\\\label{42}
B(t,r)&=&B_0(r)+\epsilon T(t)b(r),\\\label{43}
C(t,r)&=&C_0(r)+\epsilon T(t)\bar{c}(r),\\\label{44}
\rho(t,r)&=&\rho_0(r)+\epsilon {\bar{\rho}}(t,r),\\\label{45}
p_r(t,r)&=&p_{r0}(r)+\epsilon {\bar{p_r}}(t,r),\\\label{46}
p_{\perp}(t,r)&=&p_{\perp0}(r)+\epsilon{\bar{p_{\perp}}}(t,r),\\\label{47}
m(t,r)&=&m_0(r)+\epsilon {\bar{m}}(t,r),\\\label{48}
\Theta(t,r)&=&\epsilon {\bar{\Theta}}(t,r).
\end{eqnarray}
Also, the Ricci scalar in $f(R)$ model follow the same scheme as
\begin{eqnarray}\label{49'}
R(t,r)&=&R_0(r)+\epsilon T(t)e(r),\\\label{50'} f(R)&=&R_0(1+2\delta
R_0)+\epsilon T(t)e(r)(1+2\delta R_0),\\\label{51'}
F(R)&=&(1+2\delta R_0)+2\epsilon \delta T(t)e(r),
\end{eqnarray}
where $0<\epsilon\ll1$. By the freedom allowed in radial
coordinates, we choose $C_0(r)=r$ as the Schwarzschild coordinate.

The static configuration of the field equations
(\ref{13})-(\ref{16}) is obtained as
\begin{eqnarray}\nonumber
&&\frac{1}{1+2\delta R_0}\left[\kappa\rho_0-\frac{\delta
R_0^2}{2}+\frac{2\delta R_0''}{B_0^2}+\frac{2\delta
R_0'}{B_0^2}\left(\frac{2}{r}-\frac{B_0'}{B_0}\right)\right]\\\label{50}
&&=\frac{1}{(B_0r)^2}\left(2r\frac{B_0'}{B_0}+B_0^2-1\right),\\\nonumber
&&\frac{1}{1+2\delta R_0}\left[\kappa p_{r0}+\frac{\delta
R_0^2}{2}-\frac{2\delta
R_0'}{B_0^2}\left(\frac{A_0'}{A_0}+\frac{2}{r}\right)\right]\\\label{51}
&&=\frac{1}{(B_0r)^2}\left(2r\frac{A_0'}{A_0}-B_0^2+1\right),\\\nonumber
&&\frac{1}{1+2\delta R_0}\left[\kappa p_{\perp0}+\frac{\delta
R_0^2}{2}-\frac{2\delta R_0''}{B_0^2}-\frac{2\delta
R_0'}{B_0^2}\left(\frac{1}{r}+\frac{A_0'}{A_0}-\frac{B_0'}{B_0}\right)\right]\\\label{52}
&&=\frac{1}{B_0^2}\left[\frac{A_0''}{A_0}-\frac{A_0'}{A_0}\frac{B_0'}{B_0}+\frac{1}{r}
\left(\frac{A_0'}{A_0}-\frac{B_0'}{B_0}\right)\right].
\end{eqnarray}
Applying the perturbed quantities in Eqs.(\ref{41})-(\ref{51'}) to
the field equations (\ref{13})-(\ref{16}) along with
Eqs.(\ref{50})-(\ref{52}), the resulting equations
(\ref{53})-(\ref{56}) are given in \textbf{Appendix}. In its static
configuration, the dynamical equation (\ref{19}) is identically
satisfied while (\ref{21}) yields
\begin{eqnarray}\nonumber
&&p_{r0}'+(\rho_0+p_{r0})\frac{A_0'}{A_0}+\frac{2}{r}(p_{r0}-p_{\perp0})
+\frac{1}{\kappa}\left[\frac{-3R_0'}{2}-3\delta
R_0R_0'\right.\\\label{59} &&\left.+\frac{2\delta
R_0'}{B_0^2}\frac{B_0'}{B_0}\left(\frac{A_0'}{A_0}+\frac{2}{r}\right)
+\frac{4\delta R_0''}{rB_0^2}-\frac{2\delta
R_0'}{B_0^2}\frac{A_0''}{A_0}\right]=0.
\end{eqnarray}
The perturbed configurations of Eq.(\ref{19}) leads to
\begin{eqnarray}\nonumber
&&\frac{1}{A_0}\left[\dot{\bar{\rho}}+(\rho_0+p_{r0})\dot{T}\frac{b}{B_0}
+2(\rho_0+p_{\perp0})\dot{T}\frac{\bar{c}}{r}+D_3\dot{T}\right]=0,
\end{eqnarray}
where $D_3$ is given in \textbf{Appendix}. Integrating this equation
with respect to time, we get
\begin{eqnarray}\label{62}
\bar{\rho}&=&-\left[(\rho_0+p_{r0})\frac{b}{B_0}
+2(\rho_0+p_{\perp0})\frac{\bar{c}}{r}+D_3\right]T.
\end{eqnarray}
The perturbed part of Eq.(\ref{21}) is obtained as
\begin{eqnarray}\nonumber
&&\frac{bT}{B_0^2}\left[p_{r0}'+(\rho_0+p_{r0})\frac{A_0'}{A_0}
+\frac{2}{r}(p_{r0}-p_{\perp0})\right]
-(\rho_0+p_{r0})\left(\frac{a}{A_0}\right)'\frac{T}{B_0}\\\nonumber
&+&2(p_{r0}-p_{\perp0})\left(\frac{\bar{c}}{r}\right)'\frac{T}{B_0}
+\frac{\bar{p}_r'}{B_0}+(\bar{\rho}+\bar{p}_r)\frac{A_0'}{A_0}
+\frac{2}{r}(\bar{p}_r-\bar{p}_{\perp})+D_4=0
\end{eqnarray}
where a lengthy expression for $D_4$ is written in
\textbf{Appendix}. Perturbation on Eq.(\ref{18}) yields
\begin{eqnarray}\label{63}
m_0&=&\frac{r}{2}\left(1-\frac{1}{B_0^2}\right),\\\label{64}
\bar{m}&=&-\frac{T}{B_0^2}\left[r\left(\bar{c}'-\frac{b}{B_0}\right)
+(1-B_0^2)\frac{\bar{c}}{2}\right].
\end{eqnarray}
From the matching condition Eq.(\ref{30}) with Eq.(\ref{45}), it
follows that
\begin{eqnarray}\label{63'}
p_{r0}&\overset{\Sigma^{(e)}}{=}&\frac{\delta
R_0^2}{2}+\frac{2\delta
R_0'}{B_0^2}\left(\frac{A_0'}{A_0}-\frac{2}{r}\right),\\\nonumber
\bar{p}_{r}&\overset{\Sigma^{(e)}}{=}&\frac{2\delta
e\ddot{T}}{A_0^2}-\frac{2\delta \dot{T}}{A_0B_0}
\left(e'+e\frac{A_0'}{A_0}-\frac{b}{B_0}R_0'\right)+2\delta
T\left[-e\left(2\delta R_0+\frac{R_0}{1+2\delta
R_0}\right.\right.\\\nonumber &+&\left.\frac{3R_0^2}{2(1+2\delta
R_0)}\right)+\frac{R_0'}{B_0^2}
\left(\frac{a'}{A_0}+\frac{2A_0'}{A_0B_0}+\frac{2\bar{c}'}{r}-\frac{2\bar{c}}{r^2}
+\frac{4b}{B_0r}\right)\\\label{86''} &-&\left.\frac{1}{B_0^2}
\left(\frac{A_0'}{A_0}-\frac{2}{r}\right)\left(e'-\frac{\delta
eR_0'}{1+2\delta R_0}\right)\right].
\end{eqnarray}
Substituting the above equations in Eq.(\ref{55}), we obtain
\begin{equation}\label{66}
\alpha(r)\ddot{T}(t)+\beta(r)\dot{T}(t)+\gamma(r)T(t)\overset{\Sigma^{(e)}}{=}0,
\end{equation}
where $\alpha$, $\beta$, and $\gamma$ are provided in
\textbf{Appendix}. For the sake of instability region, we assume
that all the functions involved in the above equation are such that
$\alpha,~\beta$ and $\gamma$ remain positive. The corresponding
solution of Eq.(\ref{66}) is given by
\begin{equation}\label{68}
T(t)=-\exp(\omega_{\Sigma^{(e)}}t),\quad\textmd{where}\quad
\omega_{\Sigma^{(e)}}=\frac{-\beta+\sqrt{\beta^2-4\alpha\gamma}}{2\alpha}.
\end{equation}
Here we impose that the system starts collapsing at $t=-\infty$
such that $T(-\infty)=0$, keeping the system in static position.
Afterwards with the increase of $t$, it goes on collapsing. Using
Eqs.(\ref{48}) and (\ref{8}), it follows that
\begin{equation}\label{57}
\bar{\Theta}=\frac{\dot{T}}{A_0}\left(\frac{b}{B_0}+\frac{2\bar{c}}{r}\right).
\end{equation}
The expansionfree condition implies that
\begin{equation}\label{79}
\frac{b}{B_0}=-2\frac{\bar{c}}{r}.
\end{equation}
It is interesting to mention that expansionfree fluid evolution
causes a blowup of shear scalar which results the appearance of a
naked singularity \cite{H2, H4}. The dynamical instability of
collapsing fluids can be well discussed in term of adiabatic index
$\Gamma$. We relate $\bar{\rho}$ and $\bar{p_r}$ for the static
spherically symmetric configuration by assuming an equation of
state of Harrison-Wheeler type as follows \cite{ch1,HW}
\begin{equation}\label{69}
\bar{p_r}=\Gamma\frac{p_{r0}}{\rho_0+p_{r0}}\bar{\rho}.
\end{equation}
Here $\Gamma$ measures the variation of pressure for a given
variation of density. We take it constant throughout the fluid
evolution.

\section{Newtonian and Post Newtonian Terms and Dynamical Instability}

This section provides help to identify the terms belonging to
Newtonian (N), post Newtonian (pN) and post post Newtonian (ppN)
regimes. This is done by converting relativistic units into c.g.s.
units and expanding upto order $c^{-4}$ in the dynamical equation.
For the N approximation, we assume
\begin{equation}\setcounter{equation}{1}\label{70}
\rho_0\gg p_{r0},\quad\rho_0\gg p_{\perp0}.
\end{equation}
For the metric coefficients expanded upto pN approximation, we
take
\begin{equation}\label{71}
A_0=1-\frac{Gm_0}{c^2},\quad B_0=1+\frac{Gm_0}{c^2},
\end{equation}
where $G$ is the gravitational constant and $c$ is the speed of
light. From Eq.(\ref{52}), we can write
\begin{eqnarray}\nonumber
\frac{A_0''}{A_0}&=&\left(\frac{B_0'}{B_0}-\frac{1}{r}-\frac{2\delta
R_0'}{1+2\delta R_0}\right)+\frac{B_0^2}{1+2\delta
R_0}\left[\kappa p_{\perp0}+\frac{\delta
R_0^2}{2}\right]\\\label{71'} &-& \frac{2\delta}{1+2\delta
R_0}\left[R_0''+R_0'\left(\frac{1}{r}-\frac{B_0'}{B_0}\right)\right]
+\frac{1}{r}\frac{B_0'}{B_0}.
\end{eqnarray}
Also, from Eq.(\ref{63}), we obtain
\begin{equation}\label{85}
\frac{B_0'}{B_0}=\frac{-m_0}{r(r-2m_0)}.
\end{equation}
Using Eqs.(\ref{63}) and (\ref{51}), it follows that
\begin{equation}\label{72}
\frac{A_0'}{A_0}=\frac{2r^3(\kappa p_{r0}-R_0-3\delta
R_0^2)+4\delta r(R_0-2R_0'r+4R_0'm_0)+4m_0}{4r(r-2m_0)(1+2\delta
R_0+\delta R_0'r)}.
\end{equation}
Substituting Eqs.(\ref{71}), (\ref{71'}), (\ref{72}) in (\ref{59})
and doing some algebra, the first dynamical equation in relativistic
units is as follows
\begin{eqnarray}\nonumber
p_{r0}'&=&\frac{2}{r}(p_{\perp0}-p_{r0})-\left[\rho_{0}+p_{r0}+(r-2m_0)\frac{2\delta
R_0'}{r}\left(\frac{1}{r}+\frac{2\delta R_0'}{1+2\delta
R_0}\right)\right]\\\nonumber &\times&\left[ \frac{2r^3(\kappa
p_{r0}-R_0-3\delta R_0^2)+4\delta
r(R_0-2R_0'r+4R_0'm_0)+4m_0}{4r(r-2m_0)(1+2\delta R_0+\delta
R_0'r)}\right]\\\nonumber &-&\frac{2\delta R_0'}{\kappa(1+2\delta
R_0)}\left[\kappa p_{\perp0}+\frac{\delta
R_0^2}{2}\right]+\frac{2\delta (r-2m_0)R_0'}{\kappa
r^2}\\\nonumber&\times&\left[\frac{m_0}{r(r-2m_0)}\left(1+\frac{2\delta
rR_0'}{1+2\delta
R_0}\right)+2R_0''\left(\frac{1}{R_0'}+\frac{\delta r}{1+2\delta
R_0}\right)\right.\\\label{73}&+&\left.\frac{2\delta
R_0'}{(1+2\delta R_0)}\right]+\frac{3R_0'}{2\kappa}(1+2\delta
R_0).
\end{eqnarray}

In view of dimensional analysis, this equation can be written in
c.g.s. units as follows
\begin{eqnarray}\nonumber
p_{r0}'&=&-\frac{G}{c^2}
\left[\rho_{0}+c^{-2}p_{r0}+\left(\frac{r}{Gc^{-2}}-2m_0\right)\frac{2\delta
R_0'}{r}\left(\frac{1}{r}+\frac{2\delta R_0'}{1+2\delta
R_0}\right)\right]\\\nonumber &\times&\left[ \frac{2r^3(\kappa
p_{r0}-R_0-3\delta R_0^2)+4\delta
rc^4G^{-1}(R_0-2R_0'r+4R_0'Gc^{-2}m_0)+4m_0c^2}{4r(r-2Gc^{-2}m_0)(1+2\delta
R_0+\delta R_0'r)}\right]\\\nonumber &-&\frac{2\delta
R_0'c^2}{\kappa(1+2\delta R_0)}\left[\frac{\kappa
c^2p_{\perp0}}{G}+c^{-2}\frac{\delta
R_0^2}{2}\right]+\frac{2\delta (r-2m_0Gc^{-2})c^2R_0'}{\kappa
r^2}\\\nonumber&\times&\left[\frac{m_0}{r(r-2Gc^{-2}m_0)}\left(1+\frac{2\delta
rR_0'}{1+2\delta
R_0}\right)+\frac{2c^2R_0''}{G}\left(\frac{1}{R_0'}+\frac{\delta
r}{1+2\delta R_0}\right)\right.\\\label{73'}&+&\left.\frac{2\delta
R_0'c^2}{G(1+2\delta R_0)}\right]+\frac{3R_0'}{2\kappa}(1+2\delta
R_0)+\frac{2}{r}(p_{\perp0}-p_{r0}).
\end{eqnarray}
Expanding upto $c^{-4}$ order and rearranging lengthy calculations,
the resulting expression given in \textbf{Appendix}. Here the order
of $c$ differentiates the terms in (\ref{N}), (\ref{PN}) and
(\ref{PPN}) respectively as follows
\begin{eqnarray}\label{76}
&&\textmd{terms of order}~ c^0~ \textmd{correspond to
N-approximation,}
\\\label{77} &&\textmd{terms of order} ~c^{-2}~ \textmd{correspond to
pN-approximation,}\\\label{78} &&\textmd{terms of order}~ c^{-4} ~
\textmd{correspond to ppN-approximation}.
\end{eqnarray}

This strategy may help to discuss some physical results in a
certain regime by discarding terms belonging to other regimes.
Applying the expansionfree condition (\ref{79}) on (\ref{62}), we
get
\begin{eqnarray}\label{82}
\bar{\rho}&=&\left[2(p_{r0}-p_{\perp0})
\frac{\bar{c}}{r}+D_5\right]T,
\end{eqnarray}
where $D_5$ is given in \textbf{Appendix}. This equation shows that
the perturbed energy density depends on the static pressure
anisotropy and higher order corrections. This fact supports the
expansionfree condition that change in energy density depends
exclusively on pressure anisotropy. Substituting this value of
$\bar{\rho}$ in (\ref{69}), we note that the expression $\bar{p}_r$
and also $\bar{\rho}\frac{A_0'}{A_0}$ in (\ref{72}) are of ppN order
approximation. Thus we neglect them in the following calculations in
order to discuss the instability conditions upto pN order.

Applying expansionfree condition on the dynamical equation
(\ref{61}) along with value of $T$ in (\ref{68}) and following the
choice of radial functions
\begin{eqnarray}\label{a''}
a(r)=a_0r,\quad \bar{c}(r)=c_0r\quad e(r)=e_0r,
\end{eqnarray}
after tedious algebra, it follows that
\begin{eqnarray}\nonumber
&&2c_0p_{r0}'+(p_{r0}-p_{\perp0})\frac{4c_0}{r}+(\rho_0+p_{r0})\frac{a_0}{A_0}+\frac{4\delta
e_0}{\kappa r^2}+\frac{4\delta e_0}{\kappa
r}\frac{B_0'}{B_0}\\\nonumber
&-&\frac{2B_0c_0}{\kappa}\left(\frac{2r\delta
R_0}{B_0}\right)'-\frac{B_0}{\kappa}\left[\frac{2\delta
R_0'}{B_0^4}\left(\frac{a_0}{A_0}+8c_0\right)\right]_{,1}\\\nonumber
&+&\frac{2c_0B_0^2}{\kappa}\left(\frac{\delta
R_0^3}{2B_0^2}\right)'+\frac{2c_0B_0^2}{r\kappa}\left(\frac{2\delta
R_0'}{B_0^4}\right)'+\frac{2\delta R_0'}{\kappa
rB_0^2}\left(\frac{4c_0}{r}+\frac{a_0}{A_0^2}\right)\\\nonumber
&+&\frac{4e_0R_0}{\kappa}\frac{\delta
R_0^2}{2}\frac{B_0'}{B_0}+\frac{4\delta R_0'}{\kappa
B_0}e_0r(1+2\delta R_0)+\frac{8c_0B_0'}{\kappa}\\\nonumber
&-&\frac{2\delta
B_0'}{\kappa}\left[\frac{2e_0}{r}+\frac{a_0B_0R_0'}{A_0}+\frac{12c_0R_0'}{r}\right]
+\frac{32\delta c_0}{\kappa
rB_0^2}\left[\frac{R_0''}{B_0}(1-B_0)-\frac{R_0'}{r}\right]\\\nonumber
&-&\frac{4\delta e_0}{\kappa
rB_0^3}\left[\frac{1}{r}-4c_0+\frac{B_0'}{B_0}\right]-\frac{A_0'}{A_0}\left[
ra_0(\rho_0+p_{r0})+\frac{B_0}{\kappa}\left\{\frac{2\delta
R_0'}{\kappa}(4c_0\right.\right.\\\nonumber
&-&\left.a_0r)\right\}_{,1}+\frac{2\delta
e_0}{\kappa}-\frac{2c_0B_0^2}{\kappa}\left(\frac{2\delta
R_0'}{B_0^4}\right)_{,1}+\frac{2\delta}{\kappa
rA_0^2B_0^2}(a_0rR_0'+e_0+2R_0'')\\\nonumber
&+&\left.\frac{2\delta
B_0'R_0'}{\kappa}\left(\frac{e_0}{R_0}-10c_0-2c_0B_0-\frac{a_0r}{A_0}\right)
-2c_0p_{r0}\right]+\left[ \frac{4\delta c_0R_0'}{\kappa
B_0^2}\right.\\\nonumber&-&\left.\frac{2\delta R_0'}{\kappa
B_0^3}(4c_0-a_0r)-\frac{2\delta
e_0}{\kappa}\right]\left(\frac{A_0'}{A_0}\right)_{,1}-\frac{2\delta
\omega_{\Sigma^{(e)}}^2}{\kappa}\left[B_0^2\left(\frac{e_0r}{A_0^2B_0^2}\right)_{,1}\right.\\\label{92}
&+&\left.\frac{1}{A_0^2}\left\{e_0+2c_0R_0'+e_0r(A_0-1)\frac{A_0'}{A_0}\right\}\right]
+\frac{2\bar{p}_{\perp}B_0}{r} e^{-\omega_{\Sigma^{(e)}} t}=0.
\end{eqnarray}
Inserting Eqs.(\ref{71}), (\ref{85}) and (\ref{72}) upto pN order
(with $c=G=1$) in the above equation, we obtain
\begin{eqnarray}\nonumber
&&2c_0p_{r0}'+(p_{r0}-p_{\perp0})\frac{4c_0}{r}+(\rho_0+p_{r0})(r+m_0)\frac{a_0}{r}
+\frac{4\delta e_0}{\kappa r^2}+\frac{4\delta e_0m_0}{\kappa
r^3}\\\nonumber &\times&(r+2m_0)-\frac{2c_0}{\kappa r}(r+m_0)
[(r+m_0)(2\delta R_0)]_{,1}-\frac{1}{r\kappa}(r+m_0)\\\nonumber
&\times&\left[\frac{2\delta
R_0'}{r^2}(r-4m_0)(a_0(r-m_0)+8c_0r)\right]_{,1}
+\frac{2c_0}{r\kappa}(r+2m_0)\left[\frac{R_0}{2r}(r+2m_0)\right.\\\nonumber
&\times&\left.(2+3\delta
R_0)\right]_{,1}+\frac{2c_0}{r^2\kappa}(r+2m_0)\left[\frac{2\delta
R_0'}{r}(r-4m_0)\right]_{,1}+\frac{2\delta R_0'}{\kappa
r^3}(r-2m_0)
\end{eqnarray}
\begin{eqnarray}\nonumber
&\times&[4c_0+a_0(r+2m_0)]+\frac{2e_0R_0m_0}{r^3\kappa}\delta
R_0(r+2m_0)-\frac{4\delta R_0'e_0}{\kappa}\\\nonumber
&\times&(r-m_0)2\delta
R_0+\frac{8c_0m_0}{r^4\kappa}(r+2m_0)^2-\frac{2\delta
m_0^2}{r^5\kappa}(r+2m_0)^2\\\nonumber
&\times&[2e_0+\frac{a_0R_0'}{r}(r^2-m_0^2)+12c_0R_0']+\frac{32\delta
c_0}{\kappa r^4}(r-2m_0)[R_0''(r-m_0)m_0\\\nonumber
&-&R_0'r]-\frac{4\delta e_0}{\kappa
r^5}(r-3m_0)\left[r-4c_0r^2+m_0(r+2m_0) \right]\\\nonumber
&-&\left[\frac{2r^3(\kappa p_{r0}+\delta R_0^2)+4\delta
r(-2R_0'r+4R_0'm_0)+4m_0}{4r(r-2m_0)(1+2\delta R_0+\delta
R_0'r)}\right]\\\nonumber
&\times&\left[ra_0(\rho_0+p_{r0})+\frac{1}{r\kappa}(r+m_0)\left\{\frac{2\delta
R_0'}{\kappa}(4c_0-a_0r)\right\}_{,1}+\frac{2\delta
e_0}{\kappa}-\frac{2c_0}{r\kappa}\right.\\\nonumber
&\times&(r+2m_0)\left\{\frac{2\delta
R_0'}{r}(r-2m_0)\right\}_{,1}+\frac{2\delta}{\kappa
r^3}(r-2m_0)(r+2m_0)+(a_0rR_0'\\\nonumber
&+&e_0+2R_0'')+\frac{2m_0\delta
R_0'}{r^5\kappa}(r+2m_0)^2\{-10rc_0+\frac{re_0}{R_0}-2c_0(r+m_0)-a_0\\\nonumber
&\times&(r+m_0)\} -\left.2c_0p_{r0}\right]-\frac{2\delta
\omega^2}{r\kappa}\left[(r+2m_0)e_0\left(\frac{r-4m_0^2}{r}\right)_{,1}+(r+2m_0)
\right. \\\nonumber
&\times&\left.(e_0+2c_0R_0')\right]+\frac{2\bar{p}_{\perp}}{r^2}(r+m_0)
e^{-\omega_{\Sigma^{(e)}}
t}+\frac{2\delta}{r\kappa}\left.[2c_0R_0'(r+2m_0)\right.\\\nonumber
&-&\left.R_0'(r-3m_0)(4c_0-a_0r)-e_0r +\frac{2\delta
\omega_{\Sigma^{(e)}}^2e_0}{r\kappa}(r-m_0)(r+2m_0)\right]
\\\label{93} &\times&\left[\frac{2r^3(\kappa p_{r0}+\delta
R_0^2)+4\delta r(-2R_0'r+4R_0'm_0)+4m_0}{4r(r-2m_0)(1+2\delta
R_0+\delta R_0'r)}\right]_{,1}=0.
\end{eqnarray}
In general, the instability range depends upon the index $\Gamma$ as
it measures the compressibility of the fluid. However, the above
equation is independent of $\Gamma$ which shows that instability
region is totally depends upon the pressure anisotropy, energy
density, chosen $f(R)$ model and arbitrary constants. Notice that
independence of $\Gamma$ factor indicates that under expansionfree
condition, fluid evolves without being compressed. In this way, the
given $f(R)$ model shows the consistency of the physical results
with expansionfree condition.

Notice that in the above equation, some terms appearing from ppN
approximation. In the following, we are interested in discussing
the dynamic instability of N-regime, so we ignore the terms
belonging to pN and ppN approximations in Eq.(\ref{93}) as
\begin{eqnarray}\nonumber
&&2c_0|p_{r0}'|+(p_{r0}-p_{\perp0})\frac{4c_0}{r}+(\rho_0+p_{r0})(r+m_0)\frac{a_0}{r}
+\frac{4\delta e_0}{\kappa r^2}\\\nonumber &+&\frac{4\delta
e_0m_0}{\kappa r^3}(r+2m_0)-\frac{2c_0}{\kappa r}(r+m_0)
[(r+m_0)2\delta R_0]_{,1}\\\nonumber
&-&\frac{2c_0}{r\kappa}(r+2m_0)\left[\frac{R_0}{2r}(r+2m_0)\delta
R_0\right]_{,1}+\frac{2c_0}{r^2\kappa}(r+2m_0)\\\nonumber
&\times&\left[\frac{2\delta
R_0'}{r}(r-4m_0)\right]_{,1}+\frac{2\delta R_0'}{\kappa
r^3}(r-2m_0)[4c_0+a_0(r+2m_0)]\\\nonumber &-&\frac{4\delta
R_0'e_0}{\kappa}(r-m_0)(2\delta
R_0)+\frac{2\bar{p}_{\perp}}{r^2}(r+m_0) e^{-\omega_{\Sigma^{(e)}}
t}\\\nonumber &=&\frac{4\delta e_0}{\kappa
r^4}(r-3m_0)(1-4c_0r)+\frac{2\delta
\omega_{\Sigma^{(e)}}^2}{r\kappa}(r+2m_0)(e_0+2c_0R_0')\\\label{95}
&-&\frac{2e_0R_0m_0}{r^3\kappa}3\delta R_0(r+2m_0).
\end{eqnarray}
In order to fulfill the instability of expansionfree fluids, we need
to keep all the terms positive in Eq.(\ref{95}). Here, we assume
that all the arbitrary constants and dynamical quantities are
positive whereas $p_{r0}'<0$ showing that pressure decreases during
collapsing process. In addition, we need to satisfy the following
constraints
\begin{equation}\label{in1}
p_{r0}>p_{\perp0},\quad\frac{1}{4r}>c_0>0,\quad r>4m_0.
\end{equation}
Thus the system would be unstable in N-approximation as long as the
above inequalities are satisfied because other constraints on $m$
are followed by the last inequality in Eq.(\ref{in1}).

For instance, if we assume that the scalar curvature is constant,
i.e., $R_0(r)=R_c=\textmd{constant}$ and $e_0=0$, then
Eq.(\ref{95}) reduces to
\begin{eqnarray}\nonumber
&&2c_0|p_{r0}'|+(p_{r0}-p_{\perp0})\frac{4c_0}{r}+\frac{R_0c_0}{r^2\kappa}
(r+2m_0)(1+2m_0')\delta R_0\\\nonumber
&+&(r+m_0)\left[(\rho_0+p_{r0})\frac{a_0}{r}+\frac{4c_0}{\kappa
r}(1+m_0')\delta R_0+\frac{2\bar{p}_{\perp}}{r^2}
e^{-\omega_{\Sigma^{(e)}} t}\right]=0.\\\label{96}
\end{eqnarray}
Applying constant curvature condition on Eq.(\ref{24}) and
inserting in the above equation, we have
\begin{eqnarray}\nonumber
&&2c_0|p_{r0}'|+(p_{r0}-p_{\perp0})\frac{4c_0}{r}+\frac{R_cc_0}{r^2\kappa}
\left[r+rR_c+\frac{\kappa}{(1+2\delta
R_c)}\int^{r}_{\Sigma^{(i)}}\rho_0 r^2dr\right]\\\nonumber
&\times&\left[R_c+\frac{\kappa \rho_0 r^2}{1+2\delta
R_c}\right](2+3\delta
R_c)+\left[r+\frac{rR_c}{2}+\frac{\kappa}{2(1+2\delta
R_c)}\right.\\\nonumber &\times&\left.
\int^{r}_{\Sigma^{(i)}}\rho_0
r^2dr\right]\left[(\rho_0+p_{r0})\frac{a_0}{r}+\frac{2\delta
R_cc_0}{\kappa r}\left(R_c+\frac{\kappa \rho_0 r^2}{1+2\delta
R_0}\right)\right.\\\label{97}&+&\left.\frac{2\bar{p}_{\perp}}{r^2}
e^{-\omega_{\Sigma^{(e)}} t}\right]=0.
\end{eqnarray}

Let us consider an energy density profile of the form
$\rho_0=\lambda r^n$, where $\lambda$ is a positive constant and
$-\infty<n<\infty$. Substituting this value of $\rho_0$ in
Eq.(\ref{97}), it follows for $n\neq-3$
\begin{eqnarray}\nonumber
&&2c_0|p_{r0}'|+(p_{r0}-p_{\perp0})\frac{4c_0}{r}+\frac{R_cc_0}{r^2\kappa}
\left[r+rR_c+\frac{\kappa \lambda}{3(1+2\delta
R_c)}(r^{n+3}-r_{\Sigma^{(i)}}^{n+3})\right]\\\nonumber
&\times&\left[R_c+\frac{\kappa \lambda r^{n+2}}{1+2\delta
R_c}\right](2+3\delta R_c)+\left[r+\frac{rR_c}{2}+\frac{\kappa
\lambda}{6(1+2\delta R_c)}\right.\\\nonumber
&\times&\left.(r^{n+3}-r_{\Sigma^{(i)}}^{n+3})\right]\left[(\lambda
r^n+p_{r0})\frac{a_0}{r}+\frac{2\delta R_cc_0}{\kappa
r}\left(R_c+\frac{\kappa \lambda r^{n+2}}{1+2\delta
R_0}\right)\right.\\\label{97'}&+&\left.\frac{2\bar{p}_{\perp}}{r^2}
e^{-\omega_{\Sigma^{(e)}} t}\right]=0.
\end{eqnarray}
Thus instability range depends upon the positivity of
$p_{r0}-p_{\perp0}$ and $r^{n+3}-r_{\Sigma^{(i)}}^{n+3}$, i.e., the
system would hold the instability of expansionfree fluid for
$p_{r0}>p_{\perp0}$ and $r^{n+3}>r_{\Sigma^{(i)}}^{n+3}$. Finally,
it is remarked that close to the Newtonian regime, expansionfree
collapse proceeds without compression as the adiabatic index does
not involve in all the calculations.

\section{Summary}

In this paper, we have studied the problem of gravitational
collapse in $f(R)$ theory which is strongly motivated by the
observational data collected from supernova. Here the higher order
curvature terms are thought to be the origin of DE causing
acceleration which are treated as the matter part of the field
equations. We have considered the spherically symmetric collapsing
stars made up of locally anisotropic fluid and evolving under the
expansionfree condition. The curvature terms appear to affect the
passive gravitational mass and rate of collapse. Also, $f(R)$ DE
slows down the rate of collapse due to its repulsive effect.

The dynamical equations help to investigate the evolution of
gravitational collapse with time and yield the variation of total
energy inside a collapsing body with respect to time and adjacent
surfaces. We have formulated these equations by using contracted
Bianchi identities both for the usual matter and effective
energy-momentum tensor independently for a well-known
$f(R)=R+\delta R^2$ model. The first dynamical equation is used to
identify the terms belonging to Newtonian, post Newtonian and post
post Newtonian regimes. We have used the concept of relativistic
and c.g.s units. The second dynamical equation is used to discuss
the instability range of expansionfree fluid evolution upto pN
order.

Perturbation scheme is applied on the field equations and
dynamical equations. The study of resulting equations shows that
the range of instability is independent of adiabatic index
$\Gamma$ which generally plays central role in the definition of
instability range. For example, for a Newtonian perfect fluid, the
system is unstable for $\Gamma<4/3$. In our results, the
independence of $\Gamma$ shows the consistency of expansionfree
condition with $f(R)$ gravity because this condition requires that
fluid would evolve without compressibility. Moreover, the
instability range depends upon the anisotropy of radial pressure,
energy density and some constraints which arise for keeping the
positivity of the dynamical equation in Newtonian approximation.
Assumption of constant scalar curvature implies that the above
dependence of instability range also describes the instability of
cavity itself with the additional information that
$r^{n+3}>r_{\Sigma^{(i)}}^{n+3}$ should be satisfied. It is
mentioned here that at pN regime only relativistic effects are
taken into account, however, physical behavior of the dynamical
equation would be the same.

\vspace{0.5cm}

{\bf Acknowledgment}

\vspace{0.25cm}

We would like to thank the Higher Education Commission, Islamabad,
Pakistan for its financial support through the {\it Indigenous
Ph.D. 5000 Fellowship Program Batch-III}.

\section*{Appendix}

\begin{eqnarray}\setcounter{equation}{1}\nonumber
D_1&=&\frac{\dot{\rho}}{A}+\frac{\dot{B}}{AB}(\rho+p_r)+
\frac{2}{A}\frac{\dot{C}}{C}(\rho+p_{\perp})+\frac{A}{\kappa}\left[
\left\{\frac{1}{A^2B^2}\left(\dot{F}'-\frac{A'}{A}\dot{F}-
\frac{\dot{B}}{B}{F'}\right)\right\}_{,1}\right.\\\nonumber
&+&\left\{\frac{f-RF}{2A^2}+\frac{F''}{A^2B^2}-\frac{\dot{F}}{A^2}
\left(\frac{\dot{B}}{B}-\frac{2\dot{C}}{C}\right)-\frac{F'}{B^2}\left(\frac{B'}{B}
-\frac{2C'}{C}\right)\right\}_{,0}\\\nonumber
&+&\frac{\dot{A}}{A^3}\left\{\frac{f-RF}{2A^2}+\frac{F''}{A^2B^2}
-\frac{\dot{F}}{A^2}\left(\frac{\dot{B}}{B}-\frac{2\dot{C}}{C}\right)
-\frac{F'}{B^2}\left(\frac{B'}{B}-\frac{2C'}{C}\right)\right\}\\\nonumber
&+&\frac{\dot{B}}{BA^2}\left\{\frac{F''}{B^2}+\frac{\ddot{F}}{A^2}
-\frac{\dot{F}}{A^2}\left(\frac{\dot{A}}{A}+\frac{\dot{B}}{B}\right)
-\frac{F'}{B^2}\left(\frac{A'}{A}
+\frac{B'}{B}\right)\right\}\\\nonumber
&+&\frac{2\dot{C}}{CA^2}\left\{\frac{\ddot{F}}{A^2}
+\frac{\dot{F}}{A^2}\left(\frac{\dot{C}}{C}
-\frac{\dot{A}}{A}\right)-\frac{F'}{B^2}\left(\frac{A'}{A}
-\frac{C'}{C}\right)\right\} \\\label{19*}
&+&\left.\frac{1}{A^2B^2}\left(\dot{F}'-\frac{A'}{A}\dot{F}-
\frac{\dot{B}}{B}{F'}\right)\left(\frac{2A'}{A}+\frac{B'}{B}+\frac{C'}{C}\right)\right],\\\nonumber
D_2&=&\frac{p_r'}{B}+(\rho+p_r)\frac{A'}{AB}+2(p_r-p_{\perp})\frac{C'}{BC}
-\frac{B}{\kappa}\left[
\left\{\frac{1}{A^2B^2}\left(\dot{F}'-\frac{A'}{A}\dot{F}-
\frac{\dot{B}}{B}{F'}\right)\right\}_{,0}\right.\\\nonumber
&+&\left\{\frac{f-RF}{2B^2}-\frac{\ddot{F}}{A^2}+
\frac{\dot{F}}{A^2}\left(\frac{\dot{A}}{A}+\frac{2\dot{C}}{C}\right)
+\frac{F'}{B^2}\left(\frac{A'}{A}
+\frac{2C'}{C}\right)\right\}_{,1}\\\nonumber
&+&\frac{A'}{AB^2}\left\{\frac{\ddot{F}}{A^2}+\frac{F''}{B^2}
-\frac{\dot{F}}{A^2}\left(\frac{\dot{A}}{A}+\frac{\dot{B}}{B}\right)
-\frac{F'}{B^2}\left(\frac{A'}{A}-\frac{B'}{B}\right)\right\}\\\nonumber
&+&\frac{2B'}{B^3}\left\{\frac{f-RF}{2A^2}+\frac{\ddot{F}}{A^2}+\frac{\ddot{F}}{A^2}
-\frac{\dot{F}}{A^2}\left(\frac{\dot{A}}{A}+\frac{\dot{B}}{B}\right)
-\frac{F'}{B^2}\left(\frac{A'}{A}
+\frac{B'}{B}\right)\right\}\\\nonumber
&+&\frac{2\dot{C}}{CA^2}\left\{\frac{\ddot{F}}{A^2}
+\frac{\dot{F}}{A^2}\left(\frac{\dot{C}}{C}
-\frac{\dot{A}}{A}\right)-\frac{F'}{B^2}\left(\frac{A'}{A}
-\frac{C'}{C}\right)\right\}\\\label{21*}
&+&\left.\frac{1}{A^2B^2}\left(\dot{F}'-\frac{A'}{A}\dot{F}-
\frac{\dot{B}}{B}{F'}\right)\left(\frac{\dot{A}}{A}+\frac{3\dot{B}}{B}
+\frac{2\dot{C}}{C}\right)\right].
\end{eqnarray}

Perturbed field equations:
\begin{eqnarray}\nonumber
&&\frac{2T}{B_0^2}\left[\left(\frac{\bar{c}}{r}\right)''-\frac{1}{r}
\left(\frac{b}{B_0}\right)'
-\left(\frac{B_0'}{B_0}-\frac{3}{r}\right)\left(\frac{\bar{c}}{r}\right)'
-\left(\frac{b}{B_0}-\frac{\bar{c}}{r}\right)
\left(\frac{B_0}{r}\right)^2\right]\\\nonumber
&=&\frac{2Tb}{(1+2\delta R_0)B_0}\left[\kappa\rho_0-\frac{\delta
R_0^2}{2}+\frac{2\delta R_0''}{B_0^2}\right]+\frac{\kappa
\bar{\rho}}{1+2\delta R_0}\\\label{53}
&+&\frac{2T}{B_0^2}\frac{2\delta R_0'}{(1+\delta
R_0)}\left[\left(\frac{\bar{c}}{r}\right)'-\frac{1}{2}
\left(\frac{b}{B_0}\right)'\right],\\\label{54}
&&\left(\frac{\bar{c}}{r}\right)'-\frac{b}{B_0r}
-\frac{\bar{c}A_0'}{rA_0}=\frac{\delta}{1+2\delta
R_0}\left[-e'+e\frac{A_0'}{A_0}
+\frac{bR_0'}{B_0}\right],\\\nonumber
&-&\frac{2\ddot{T}}{A_0^2}\frac{\bar{c}}{r}+
\frac{2T}{rB_0^2}\left[\left(\frac{a}{A_0}\right)'+
\left(r\frac{A_0'}{A_0}+1\right)\left(\frac{\bar{c}}{r}\right)'-\frac{B_0^2}{r}
\left(\frac{b}{B_0}-\frac{\bar{c}}{r}\right)\right]\\\nonumber
&=&\frac{2Tb}{(1+2\delta R_0)B_0}\left(\kappa p_{r0}+\frac{\delta
R_0^2}{2}\right)+\frac{\kappa \bar{p_r}}{1+2\delta
R_0}+\frac{\ddot{T}}{A_0^2}\frac{2\delta e}{(1+2\delta
R_0)}\\\label{55} &+&\frac{2T}{rB_0^2}\frac{2\delta
R_0'}{(1+\delta
R_0)}\left[\left(\frac{a}{A_0}\right)'+2\left(\frac{\bar{c}}{r}\right)'\right],\\\nonumber
&-&\frac{\ddot{T}}{A_0^2}\left[\frac{b}{B_0}+\frac{\bar{c}}{r}\right]+\frac{T}{B_0^2}
\left[\left(\frac{a}{A_0}\right)''
+\left(\frac{\bar{c}}{r}\right)''+\left(\frac{2A_0'}{A_0}-\frac{B_0'}{B_0}+\frac{1}{r}\right)
\left(\frac{a}{A_0}\right)'\right.\\\nonumber &-&\left.
\left(\frac{A_0'}{A_0}+\frac{1}{r}\right)\left(\frac{b}{B_0}\right)'
+\left(\frac{A_0'}{A_0}-\frac{B_0'}{B_0}+\frac{2}{r}\right)
\left(\frac{\bar{c}}{r}\right)'\right]=\frac{\ddot{T}}{A_0^2}\frac{2\delta
e}{1+2\delta R_0}\\\nonumber
&+&\frac{\kappa\bar{p}_{\perp}}{1+2\delta
R_0}-\frac{T}{B_0^2}\frac{2\delta R_0'}{(1+2\delta
R_0)}\left[\left(\frac{a}{A_0}\right)'-\left(\frac{b}{B_0}\right)'
+\left(\frac{\bar{c}}{r}\right)' \right]\\\label{56}
&-&\frac{2T}{B_0(1+2\delta R_0)}\left(\kappa
p_{\perp0}+\frac{\delta R_0^2}{2}\right).
\end{eqnarray}

\begin{eqnarray}\nonumber
D_3&=&\frac{1}{A_0\kappa} \left[\frac{3a}{A_0}\frac{\delta
R_0^2}{2}-2\delta e\left(2\delta
R_0+\frac{A_0'}{r^2A_0B_0^2}\right)-\frac{2\delta }{\kappa
B_0^2}\left(e'+e\frac{A_0'}{A_0}\right)\left(\frac{A_0'}{A_0}-
\frac{1}{r}\right)\right.\\\nonumber &&+\frac{2\delta
R_0''}{B_0^2}\left(\frac{3a}{A_0}-\frac{2b}{B_0}\right)+
\frac{2\delta
R_0'}{B_0^2}\left\{\frac{2\bar{c}'}{r}+\frac{a}{A_0}\frac{B_0'}{B_0}
-\frac{b'}{B_0}+\frac{2b}{B_0}\frac{B_0'}{B_0}
-\left(\frac{b}{B_0}\right)'\right.\\\label{60}&&\left.\left.
-\frac{b}{B_0}+\frac{2a}{rA_0}+\frac{4\bar{c}}{r}\left(\frac{1}{r}
-\frac{A_0'}{A_0}\right)\right\}\right].
\end{eqnarray}

\begin{eqnarray}\nonumber
D_4&=&-\frac{B_0\ddot{T}}{\kappa}\left(\frac{2\delta
e}{A_0^2B_0^2}\right)'-\frac{2\delta \ddot{T}}{\kappa
B_0A_0^2}\left(e'-\frac{A_0'e}{A_0}-\frac{R_0'b}{B_0}\right)
-\frac{T}{\kappa}\left(\frac{2\delta
R_0e}{B_0^2}\right)'\\\nonumber
&+&\frac{T}{\kappa}\left[\frac{2\delta
R_0'}{B_0^4}\left\{\left(\frac{a}{A_0}\right)'-\left(\frac{\bar{c}}{r}\right)'
-\frac{2b}{B_0}\frac{A_0'}{A_0}-\frac{4b}{rB_0}\right\}+\frac{2\delta
e'}{B_0^2}\left(\frac{A_0'}{A_0}+\frac{2}{r}\right)\right]_{,1}\\\nonumber
&+&\frac{Tb}{\kappa^2}\left[\frac{2\delta
R_0^2}{2B_0^2}+\frac{2\delta
R_0'}{B_0^4}\left(\frac{A_0'}{A_0}+\frac{1}{r}\right)\right]_{,1}-
\frac{2\delta e\ddot{T}}{A_0B_0\kappa}\frac{A'_0}{A_0} \\\nonumber
&-&\frac{Tb}{A_0B_0^2\kappa}\left[\frac{2\delta
R_0''A_0'}{B_0^2}-\frac{2\delta
R_0'}{rA_0B_0^3}\left(\frac{A_0'}{A_0}+\frac{2}{r}\right)\right]
+\frac{T}{B_0^2\kappa}\left[\frac{\bar{c}''
A_0'}{A_0B_0}-\frac{4\delta
R_0''}{B_0}\frac{b}{B_0}\frac{A_0'}{A_0}\right.\\\nonumber
&+&\frac{2\delta
R_0''}{A_0B_0}\left(a'-a\frac{A_0'}{A_0}-2\frac{A_0'b}{B_0}\right)
-\frac{1}{A_0^2B_0}\left(2\delta R_0'a-(2\delta
R_0'a-e')\frac{A_0'}{A_0}\right.\\\nonumber &-&\left.\frac{4\delta
R_0'bA_0'}{A_0B_0}\right)-\frac{4\delta
R_0'}{rB_0}\left(\bar{c}'-\frac{\bar{c}}{r}\right)-\delta
R_0^2\left(b'-\frac{3bB_0'}{B_0}\right)\\\nonumber &+&4\delta
B_0'e(1+2\delta R_0)+2\delta
e'B_0B_0'\left(\frac{A_0'}{A_0}+\frac{2}{r}\right)
+\frac{2B_0}{r}\left(b'-\frac{3bB_0'}{B_0}\right)\\\nonumber
&+&\frac{4\delta}{rB_0^2}\left(R_0''-\frac{R_0'}{r}\right)
\left(\bar{c}'-\frac{\bar{c}}{r}-\frac{4b}{B_0}\right)
-\frac{8\delta
R_0'}{r^2B_0^2}\left(\bar{c}'-\frac{\bar{c}}{r}-\frac{4b}{B_0}\right)+\frac{4\delta
e''}{rB_0^2}\\\nonumber &+&2\delta
R_0'B_0'B_0\left\{\frac{A_0'}{A_0}\left(b-a-\frac{5b}{B_0}\right)
+\frac{a'B_0}{A_0}+\frac{2\bar{c}'}{r} -\frac{2\bar{c}}{r^3}
-\frac{4b}{B_0r}\right\}
\\\label{61} &-& \left.\frac{4\delta}{rB_0^2}\left(\frac{e'B_0'}{B_0}
+\frac{R_0'b'}{B_0}-\frac{R_0'b}{B_0r}\right) +\frac{4\delta
e'}{r^2B_0^2} \right].
\end{eqnarray}

\begin{eqnarray}\label{63''}
\alpha(r)&\overset{\Sigma^{(e)}}{=}&\frac{1}{A_0^2}\left(\frac{2\bar{c}}{r}+
\frac{\kappa e+e}{1+2\delta R_0}\right),\\
\beta(r)&\overset{\Sigma^{(e)}}{=}&\frac{2\delta
\kappa}{A_0B_0(1+2\delta
R_0)}\left(e'+\frac{eA_0'}{A_0}-\frac{bR_0'}{B_0}\right),\\\nonumber
\gamma(r)&\overset{\Sigma^{(e)}}{=}&\frac{\kappa }{B_0^2(1+2\delta
R_0)}\left[2\delta e B_0^2\left\{-2\delta R_0+\frac{1}{(1+2\delta
R_0)}\frac{\delta R_0^2}{2}\right\}\right.\\\nonumber
&+&2\delta\left(\frac{A_0'}{A_0}-\frac{2}{r}\right)\left(e'-\frac{2\delta
eR_0'}{1+2 \delta R_0}\right)+\frac{2\delta R_0'}{\kappa}
\left(\frac{a'}{A_0}+\frac{2A_0'}{A_0B_0}+\frac{2\bar{c}'}{r}\right.\\\nonumber
&-&\left.\frac{2\bar{c}}{r^2}+\frac{4b}{B_0r}\right)+\frac{4\delta
R_0}{\kappa}\left\{\left(\frac{a}{A_0}\right)'+
\left(r\frac{A_0'}{A_0}+1\right)\left(\frac{\bar{c}}{r}\right)'-\frac{B_0^2}{r}
\left(\frac{b}{B_0}-\frac{\bar{c}}{r}\right)\right\}
\end{eqnarray}
\begin{eqnarray}\nonumber
&+&2bB_0\left\{\left(1+\frac{1}{\kappa}\right)\frac{\delta
R_0^2}{2} -\frac{2\delta
R_0'}{B_0^2}\left(\frac{A_0'}{A_0}-\frac{2}{r}\right)\right\}\\&+&\left.\frac{4
\delta R_0'}{\kappa}\left\{\left(\frac{a}{A_0}\right)'+2
\left(\frac{\bar{c}}{r}\right)'\right\}\right].
\end{eqnarray}

\begin{eqnarray}\nonumber
p_{r0}'&=&\frac{2}{r}(p_{\perp0}-p_{r0})+\frac{3R_0'}{2\kappa}(1+2\delta
R_0)-\frac{\delta^2 R_0'R_0^2}{\kappa}(1-2\delta R_0)\\\nonumber
&-&\frac{4\delta R_0'm_0^2G}{\kappa r^4}(1+2\delta R_0'r(1-2\delta
R_0))+\frac{4\delta^2 R_0'^2 G}{r \kappa G}(1-2\delta R_0)
\\\nonumber &+&\frac{4\delta R_0''}{r\kappa G}(1+\delta R_0'r(1-2\delta
R_0))-\delta R_0'(\kappa p_{r0}-R_0-3\delta R_0^2)(1+2\delta
R_0'r\\\nonumber &\times&(1-2\delta R_0))-\frac{4\delta
R_0'Gm_0^2}{r^4}(1-2\delta R_0-\delta R_0'r)(2G\delta R_0+1)(1
+2r\delta R_0'\\\label{N} &\times&(1+2\delta R_0')) \\\nonumber
&-&\frac{G}{c^{2}r^3}\left[\left\{\frac{2\delta R_0'm_0}{\kappa
G}+\frac{8G\delta R_0'm_0^3}{\kappa r^2}\right\}(1+4\delta^2
R_0'R_0r)\right.\\\nonumber &+&\left\{\frac{r^3}{2}\left(\rho_0
r-\frac{2\delta R_0'Gm_0}{r}+8\delta^3 R_0'^2 R_0m_0\right)(\kappa
p_{r0}+\delta R_0^2)\right.\\\nonumber&-&2m_0\delta R_0'r^2(\kappa
p_{r0}+\delta R_0^2)(1+2r\delta^2 R_0'R_0)\\\nonumber &+&
\frac{2Gm_0^2}{r^2}(1+GR_0\delta)(\rho_0 r^2-2\delta
R_0'm_0+4m_0\delta^2 R_0'^2(1-2\delta
R_0))\\\label{PN}&+&\left.\left.\frac{8G\delta
R_0'm_0^2}{r}(1+4\delta m_0 R_0')(1+2\delta
rR_0')\right\}(1-2\delta R_0-\delta R_0'r)\right]\\\nonumber
&-&\frac{G}{c^{4}r^4}\left[\frac{16\delta R_0'm_0^4}{\kappa
r^2}(1+2\delta^2 R_0'rR_0)+\left\{r^5p_{r0}(\kappa
p_{r0}-R_0-3\delta R_0^2)\right.\right.\\\nonumber
&+&r^3Gm_0(\kappa p_{r0}-R_0-3\delta R_0^2)\left(G\rho_0
r-\frac{2\delta R_0'm_0}{r}+4\delta^3
R_0'^2R_0m_0\right)\\\nonumber&+&4Gm_0^2\delta R_0'r^2(\kappa
p_{r0}+\delta R_0^2)(1+2r\delta^2
R_0'R_0)+2rGm_0^2p_{r0}(1+2GR_0\delta)\\\nonumber &+&
\frac{8\delta GR_0'm_0^3}{r}(1+4r\delta^2 R_0')(1+2r\delta^2
R_0'R_0)+4G^2m_0^3(1+4r\delta
R_0')\\\label{PPN}&\times&\left.\left.\left(\rho_0-\frac{2\delta
R_0'm_0}{r^2}+\frac{4\delta^3R_0'R_0^2m_0}{r}\right)\right\}(1-2\delta
R_0-\delta R_0'r)\right].
\end{eqnarray}

\begin{eqnarray}\nonumber
D_5&=&\frac{3a}{\kappa A_0}\frac{\delta R_0^2}{2}-\frac{4\delta
e''}{\kappa B_0^2}+\frac{2\delta e}{\kappa}\left(2\delta
R_0+\frac{A_0'}{r^2A_0B_0^2}\right)+\frac{2\delta }{\kappa
B_0^2}\left(e'+e\frac{A_0'}{A_0}\right)\\\nonumber
&\times&\left(\frac{A_0'}{A_0}-\frac{1}{r}\right) -\frac{2\delta
R_0''}{B_0^2}\left(\frac{3a}{A_0}+\frac{4\bar{c}}{r}\right)-
\frac{2\delta R_0'}{\kappa
B_0^2}\left\{\frac{2B_0\bar{c}'}{r}+\frac{a}{A_0}\frac{B_0'}{B_0}\right.\\\label{83}
&+&\left.\frac{3\bar{c}'}{r}+\frac{3\bar{c}}{r^2}
+\frac{4\bar{c}}{r}\left(\frac{B_0'}{B_0}-\frac{A_0'}{A_0}\right)\right\}.
\end{eqnarray}

\end{document}